\documentclass{aa}
\usepackage{graphics,epsf,epsfig,graphicx}
\usepackage{natbib}
\usepackage{url}
\usepackage{pstricks,psfrag}
\usepackage{amsmath,amssymb,amsfonts}

\bibpunct{(}{)}{;}{a}{}{,} 

\newcommand{\be}{\begin{equation}}
\newcommand{\ee}{\end{equation}}
\newcommand{\dd}{\mathrm{d}}

\newcommand{\Ncodes}{seven}
\newcommand{\NMCcodes}{four}

\def\ProDiMo{{\sf ProDiMo}}
\def\mcfost{{\sf MCFOST}}
\def\mc3d{{\sf MC3D}}
\def\mcmax{{\sf MCMax}}
\def\radmc{{\sf RADMC}}
\def\raytrace{{\sf RAYTRACE}}
\def\radical{{\sf RADICAL}}
\def\radicalvet{{\sf RADICAL-VET}}
\def\torus{{\sf TORUS}}
\def\pinball{{\sf Pinball}}

\def\Td{T_{\hspace*{-0.2ex}\rm d}}

\def\r0{\vec{r}_0}

\newcommand{\vect}[1]{\overrightarrow{#1}}

\begin{document}

\title{Benchmark problems for continuum radiative transfer}
\subtitle{High optical depths, anisotropic scattering, and polarisation}

\author{C.~Pinte\inst{1}, T.~J.~Harries\inst{1}, 
M.~Min\inst{2}, A.~M.~Watson\inst{3}, C.~P.~Dullemond\inst{4},
P.~Woitke\inst{5,6}, F.~M\'enard\inst{7}, M.~C.~Dur\'an-Rojas\inst{3}}

\offprints{C. Pinte \\ \email{pinte@astro.ex.ac.uk}}
\institute{School of Physics, University of Exeter, Stocker Road,
  Exeter EX4 4QL, United Kingdom
\and
Sterrenkundig Instituut Anton Pannekoek, Universiteit Amsterdam,
Kruislaan 403, 1098 Amsterdam, The Netherlands
\and
Centro de Radioastronom\'ia y Astrofs\'ica, Universidad 
Nacional Aut\'oma de M\'exico, Morelia, Mich., M\'exico.
\and
 Max Planck Institute for Astronomy, K\"onigstuhl 17, 69117
 Heidelberg, Germany
\and
UK Astronomy Technology Centre, Royal Observatory, Edinburgh,
Blackford Hill, Edinburgh EH9 3HJ, UK
\and
School of Physics \& Astronomy, University of St.~Andrews,
             North Haugh, St.~Andrews KY16 9SS, UK 
\and
Laboratoire d'Astrophysique de Grenoble, CNRS/UJF UMR~5571, 
414 rue de la Piscine, B.P. 53, F-38041 Grenoble Cedex 9, France
}

\authorrunning{C. Pinte et al.}
\titlerunning{Benchmark problems for continuum radiative transfer}

\date{Received ... / Accepted ...}

\abstract{}{Solving the continuum radiative transfer equation in high opacity
  media requires sophisticated numerical tools. In order to test the reliability of such
  tools, we present a benchmark of radiative transfer
  codes in a 2D disc configuration.}
{We test the accuracy of \Ncodes\ independently developed radiative transfer codes by comparing the
  temperature structures, spectral energy distributions, scattered
  light images, and linear polarisation maps that each model predicts for a
  variety of
  disc opacities and viewing angles. The test cases have been chosen
  to be numerically challenging, with midplane optical depths up
  $10^6$, a sharp density transition at the inner edge and complex
  scattering matrices.  We also review recent progress in the
  implementation of the Monte Carlo method that allow an efficient
  solution to these kinds 
  of problems and discuss the advantages and limitations of Monte
  Carlo codes compared to
  those of discrete ordinate codes.  
}{For each of the test cases, the predicted results from the radiative transfer codes
  are within good
  agreement. 
The results indicate that these codes can be
  confidently used to interpret  present and future
  observations of protoplanetary discs.}{}

\maketitle

\section{Introduction}
Dust represents an essential element in the energy balance of a
variety of astrophysical objects, from the
interstellar medium to the atmospheres and close circumstellar
environments of numerous classes of object; from the lower mass
planets and brown dwarfs, to massive stars. With the advent of high-angular resolution and high-contrast instruments, the basic structural
properties (e.g., size, inclination, and surface brightness) of the
circumstellar environments of the nearest and/or largest objects ---
discs and envelopes around young stars in nearby star-forming
regions and around more distant evolved stars --- are now under close
scrutiny. 
With this unprecedented wealth of high-resolution data,
from optical to radio, detailed studies of the dust content become
possible and sophisticated radiative transfer (RT) codes are needed to fully
exploit the data.

At short wavelengths, dust grains efficiently absorb, scatter, and
polarise the starlight while at longer wavelengths dust re-emits the
absorbed radiation. How much radiation is scattered and absorbed is a
function of both the geometry of the circumstellar environment and the
properties of the dust. In turn, the amount of absorbed radiation
sets the temperature of the dust (and gas) and defines the amount of
radiation that is re-emitted at longer, thermal wavelengths.

To get a reliable understanding of the structure and evolution of
these ``dusty'' objects, be it the evolution of dust grain sizes, the
temperature dependent chemistry, or simply the density profiles, it is
highly desirable to model not only the integrated fluxes (\textit{i.e.}, the
spectral energy distributions, hereafter SED), but also the resolved
brightness
and/or polarisation profiles when available.  This can only be done by
solving the radiative transfer (hereafter RT) problem in media that
can have large optical depths and/or complex geometries and
compositions. 
Recent studies of circumstellar discs are based on detailed comparisons of high-quality
data sets, combining various kinds of observation (SED, multiple
wavelength scattered light images, polarisation map, infrared or
millimetre visibilities) to the predictions of
RT codes
\citep[e.g.][]{Wood02,Watson04b,Wolf03c,D'Alessio06,Steinacker06,Doucet07,Fitzgerald07,Pontoppidan07,Pinte07,Pinte08,Pinte08b,Glauser08,Tannirkulam08b}.
Such studies will become more and more common with the advent of
new instruments like \emph{VLT/SPHERE}, \emph{Gemini/GPI}, \emph{JWST},
\emph{Herschel} and \emph{ALMA}, and validating 
the accuracy of RT codes is of particular importance.

\defcitealias{Pascucci04}{P04}
\def\P04{\citetalias{Pascucci04}}

Analytical solutions do not  exist for wavelength-dependent
radiative transfer and sophisticated numerical methods must be used. 
Testing the reliability of RT computations requires in that case to compare the
solutions to well-defined problems by independent codes. Such a work
has been done by \cite{Ivezic97} for a 1D spherical geometry and by 
\citet[][hereafter P04]{Pascucci04} in a 2D disc configuration. The
later work compared in detail the calculations of five radiative codes
and has been used as a reference to validate newly developed RT
codes \citep[e.g.][]{Harries04,Ercolano05,Pinte06}. 
 The test cases were however limited to relatively modest optical depths
(midplane opacity $\tau < 100$ in the $V$ band), orders of magnitude smaller
than the actual optical depths of protoplanetary discs for which
radiative transfer codes are generally used in the literature. 
High optical depths represent challenging calculations for radiative transfer
codes, with potential convergence issues, and additional tests are
required to confidently trust the results of radiative codes in this regime. 
Furthermore, calculations in \P04 were
done assuming isotropic scattering and restricted to SEDs. With the
advent of high-resolution observations of young stellar objects,
validating the calculations of resolved surface brightness of these
objects is now also crucially needed.

In this
paper, we extend the work of \P04 to realistic, very optically thick
discs with anisotropic scattering.
We perform a comparison of \Ncodes\ RT codes in a well defined
2-dimensional disc configuration, with simple dust properties. 
The prediction of \NMCcodes\ Monte Carlo codes (temperature structure in the disc, the
emergent SED as well as monochromatic scattered light images, and
polarisation maps for different disc opacities and viewing angles) are
compared in the case of anisotropic scattering (sections~\ref{sec:RT}
to \ref{sec:results}). Additional comparisons with discrete ordinate codes, 
in the case of isotropic scattering and without scattering, are presented in
section~\ref{sec:grid-based}.

\section{Radiative transfer modelling}
\label{sec:RT}

\subsection{The radiative transfer problem}

Solving the RT problem in dusty environments aims at determining
the (polarised) specific intensity $\vec{I}_\lambda(\vect{r},\vect{n})$
at each point $\vect{r}$ and direction $\vect{n}$ of the volume and at
each wavelength $\lambda$. This intensity is obtained by solving the 
stationary transfer equation. 

In the case of randomly oriented dust particles, the radiative
transfer equation can be written adopting the Stokes
formalism:

\begin{multline}
\frac{\dd \vec{I}_\lambda(\vect{r},\vect{n})}{\dd s} =  -\kappa^{\mathrm{ext}}_\lambda(\vect{r})\,
\vec{I}_\lambda(\vect{r},\vect{n})\\
 + \kappa^{\mathrm{abs}}_\lambda(\vect{r})\, B_\lambda(T(\vect{r}))\,\vec{I}_0 \\
 + \kappa^{\mathrm{sca}}_\lambda(\vect{r})\frac{1}{4\pi}\,
\iint_\Omega \vec{S}_\lambda(\vect{r},\vect{n}',\vect{n})\,  \vec{I}_\lambda(\vect{r},\vect{n}') \,\dd\Omega'
\label{eq:transfert_radiatif}
\end{multline}
where $\vec{I}_\lambda(\vect{r},\vect{n}) = (I,Q,U,V)$ is the Stokes
vector, with $I$ representing the total intensity, $Q$
and $U$ the linearly polarised intensities, and $V$ the circularly
polarised intensity. 
$\kappa^{\mathrm{abs}}_\lambda(\vect{r})$,
$\kappa^{\mathrm{sca}}_\lambda(\vect{r})$ and
$\kappa^{\mathrm{ext}}_\lambda(\vect{r}) =
\kappa^{\mathrm{abs}}_\lambda(\vect{r}) +
\kappa^{\mathrm{sca}}_\lambda(\vect{r})$ are the absorption,
scattering and extinction opacities, respectively.
$s$ is the length along the direction of
propagation. $\vec{S}_\lambda(\vect{r},\vect{n}',\vect{n})$ is the
$4\times4$ scattering
(or Mueller) matrix  describing the changes in the Stokes vector when
the light is 
scattered from the direction $\vect{n}'$ to the direction $\vect{n}$.
 $B_\lambda(T)$ is the Planck function and $\vec{I}_0$ is the unitary Stokes
 vector representing unpolarised emission  $\vec{I}_0 =
 (1,0,0,0)$\footnote{because the grains are randomly oriented, the
   nett dust thermal emission is not polarised.}.

Computation of the thermal emission requires to determine the
dust temperature structure $T(\vect{r})$. This temperature is
determined by writing that the dust is in radiative equilibrium. If
we assume that the dust is at the local thermodynamic equilibrium and that
there is no more sources of energy than the radiation field, the
temperature is obtained by solving the implicit equation:
\begin{equation}
\int_0^\infty \kappa^\mathrm{abs}_\lambda(\vect{r})\, B_\lambda(T(\vect{r}))\,\dd\lambda = 
\int_0^\infty \kappa^\mathrm{abs}_\lambda(\vect{r})\, J_\lambda\,\dd\lambda
\label{eq:eq_rad}
\end{equation}
where $J_\lambda$ is the mean specific intensity (\emph{i.e.} the
specific intensity averaged over all solid angles).

The system of equations \ref{eq:transfert_radiatif} and
\ref{eq:eq_rad} completely defines the RT problem when the dust optical properties
($\kappa^{\mathrm{abs}}_\lambda$, $\kappa^{\mathrm{sca}}_\lambda$,
$\vec{S}_\lambda$) and sources of radiation (initial conditions for
equation \ref{eq:transfert_radiatif}) are given. 
It is important to note that opacities can depend on the
temperature, in which case solving simultaneous equations  \ref{eq:transfert_radiatif} and
\ref{eq:eq_rad} requires an iterative scheme. Most of the time however,
dust opacities do not vary much with temperature and can be assumed to
be constant. We will make this assumption in the
following analysis.

Additionally, radiative transfer plays an
integral role in the physics of the disc. It alters the density structure
via hydrostatic equilibrium \citep[e.g.][]{Walker04} and
impacts the dust content, and hence the opacity. For instance differential
dust sublimation at the inner edge can destroy some of the dust
grains, the temperature of a dust grain depending on its 
size \citep[e.g.,][]{Tannirkulam07} and composition
\citep[e.g.,][]{Woitke06}. Similarly the formation of ice mantles around the
grains in the cold,  outer regions of the disc also affects the grain
opacities. Including any of these effects requires an iterative approach. In this
paper, we restrict ourselves to the benchmark of radiative
transfer solvers and keep the density structure and dust properties fixed.

\begin{figure*}
  \includegraphics[height=0.37\hsize]{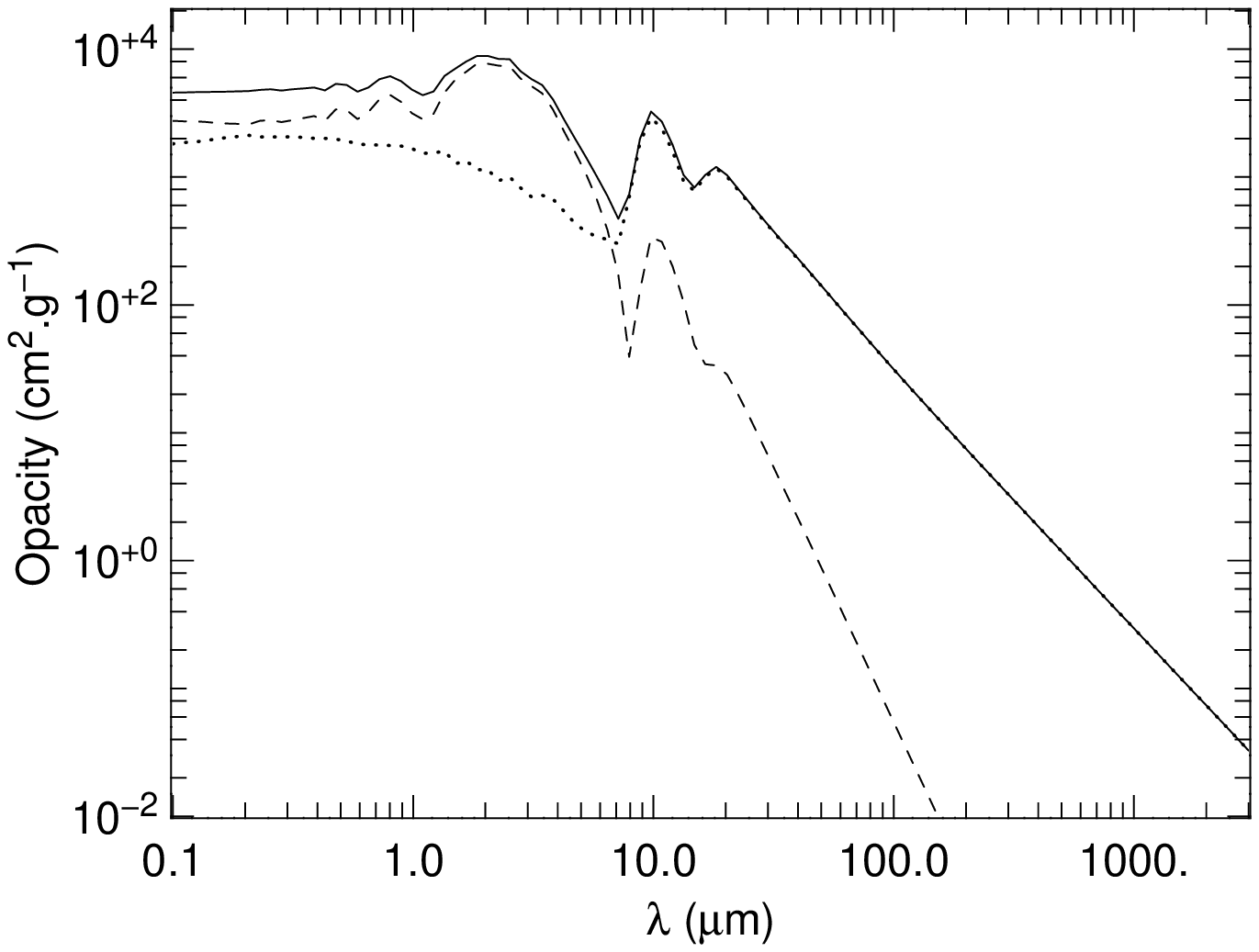}
  \hfill
  \includegraphics[height=0.37\hsize]{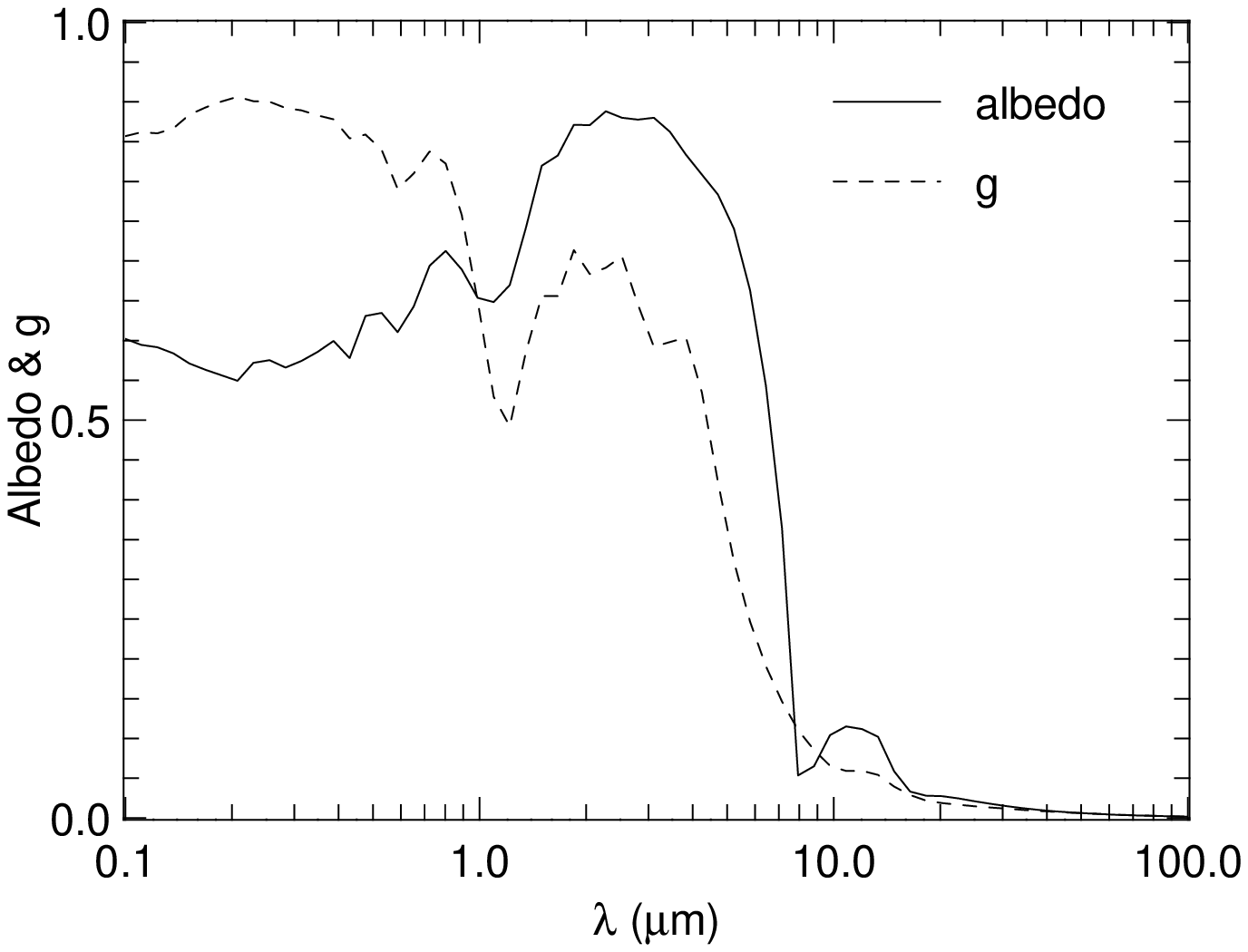}
  \caption{Dust optical properties for the 1\,$\mu$m silicate grains used in the
    calculations. \emph{Left:} dust opacity, the full line represents the extinction opacity,
    the dashed line the absorption opacity and the dotted line the
    scattering opacity. \emph{Right:} albedo (full line) and asymmetry
    parameter $g = \left<\cos \theta \right>$ (dashed line). For wavelengths
    larger than 100\,$\mu$m, both the albedo and asymmetry parameter
    have values close to $0$ and the contribution from scattered
    light is negligible. \label{fig:opacite}}
\end{figure*}

\subsection{Numerical methods}

\subsubsection{The Monte Carlo method}

Anisotropic scattering by dust grains precludes the use of
direct methods to solve the continuum RT equation and 
 Monte Carlo methods are commonly used instead. They solve the RT equation by stochastically 
propagating  ``photon packets'' through the dusty environment.
 The transport of packets is governed by scattering, absorption and re-emission events
that are controlled by the optical properties of the medium (opacity,
albedo, scattering phase function, etc) and by the temperature
distribution. Upon leaving the model boundaries, ``photon
packets'' are used to build an SED and/or synthetic images. 

The Monte Carlo scheme estimates physical quantities by statistical
means, which potentially leads to noisy results when the number of
packets sampling some regions and/or directions in the model becomes low.
Several variance reduction techniques have been developed to improve the
sampling of the Monte Carlo method:  forced first
scattering \citep{Cashwell59}, peel-off techniques \citep{Yusef84},
estimation of the mean specific intensity 
\citep{Lucy99}, immediate re-emission \citep{Bjorkman01}, and importance
weighting schemes \citep{Juvela05}. These various techniques allowed
the Monte Carlo method to progressively become competitive against
grid-based methods, and it is now more and more commonly used to solve
the continuum RT problem.

Despite these techniques, Monte Carlo methods can become
computationally expensive when the optical depth becomes very large.   
In our disc configuration, this leads to two major difficulties:
\begin{itemize}
\item because gradients of opacity are oriented toward the disc
  surface, packets entering the disc tend to escape after a few
  interactions. Very few packets penetrate the central regions of the disc,
  leading to a noisy temperature structure close to the disc midplane.
\item for edge-on configurations, the flux in the 
  near- and mid-infrared is dominated by packets originating from
  the dense central regions of the disc. These packets will be scattered towards
  the observer in the surface layers of the disc, where the probability of
  interaction is very low (due the very low density). Packets escaping
  the dense regions (after a large number of interactions) almost
  always go through the surface layers without 
  interacting with the dust grains and large number of packets are required to
  converge SEDs or images in this regime.
\end{itemize}

In the next paragraphs, we describe two schemes that, when
coupled with a Monte Carlo approach, significantly improve the
efficiency of the codes. They have helped
to overcome the previously mentioned difficulties and to efficiently
solve the test cases presented in this paper.

\subsubsection{Diffusion approximation}

In the deep regions of the disc, solving the complete RT equation is
not necessary. The radiation field becomes isotropic
and the source function becomes equal to the Planck function.
The behaviour of the density 
of energy $\epsilon(\vect{r}) = 4 \sigma T(\vect{r})^4/ c$ 
is in that case properly
described by the diffusion theory:
\begin{equation}
  \nabla . \left( D(\vect{r}) \nabla \epsilon(\vect{r}) \right) = 0
\label{eq:diff_approx}
\end{equation}
where the diffusion coefficient is defined as $D(\vect{r}) =
1/3\rho(\vect{r}) \kappa_R(\vect{r})$ with $\kappa_R(\vect{r})$
 the local Rosseland opacity.

In this case, Monte Carlo methods can be efficiently coupled to diffusion
approximation methods. The model can be divided into two regions. The
first one, that corresponds the surface of the disc represents all
parts of the model volume where the optical depth in any direction is
smaller than a given threshold. In this region, the temperature
structure is computed with a Monte Carlo method, eventually including
acceleration schemes such as the immediate re-emission concept and/or
estimation of the mean specific intensity.

In the rest of the model volume, \emph{i.e.} in the central regions of
the disc, the temperature structure can be solved using the diffusion
approximation. The temperature structure at the edge of the diffusion
approximation region (initial condition for equation~\ref{eq:diff_approx}) is
given by the solution of the Monte Carlo calculations. 

The optical depth threshold which defines these two regions must be set
high enough to ensure that the radiation field inside the diffusion
approximation region is isotropic and dominated by the local emission.

For an optimal efficiency, this method must avoid calculating the
complete propagation of photon packets inside the diffusion
approximation region. This can be done by using various methods. The
propagation of the packets can be calculated in a faster way by using a
modified random walk procedure \citep{Fleck84,Min09}. This
method combines multiple interaction steps in one computation, while
still keeping track of the energy deposited in an accurate sense. In
this way, the interaction between the optically thick regions and the
upper layers of the disc is properly computed. In addition, the number
of steps taken can be easily adjusted to the local radiation field and
density gradient, making it a highly flexible method. 
A ``mirror'' condition can also be used: when a packet enters the
diffusion approximation region, it is sent back with the same energy
and wavelength but with an opposite direction vector. Although not
rigorously exact, this method (used by \torus\ and \mcfost) was found to
be very accurate when compared to the full Monte Carlo solution
(see section~\ref{seq:temperature}).

Details and tests on the accuracy of diffusion approximation methods
are presented in \cite{Min09}.

\subsubsection{Ray-tracing}
When the specific intensity is known in
each point of the model, direct ray-tracing
methods using the formal solution of the RT equation can be used to
produce observables. 

Ray-tracing has for instance been used in combination with Monte Carlo methods to produce
SEDs and emission maps in the infrared and millimetre regimes, where scattering can be
considered isotropic in some cases \citep{Wolf03,Dullemond04a}. When
the scattering is isotropic, only the mean specific intensity and
temperature structure are
required to calculate the source function. These two quantities can be
easily estimated with a Monte Carlo method \citep{Lucy99} and do not
require large amount of memory to be stored.
After an initial Monte Carlo run computing the total mean intensity
and temperature structure 
in the disc, SEDs and/or maps can then be produced by integrating the source function on
rays originating from the observer. 
Which such a method, the Monte Carlo method is used only to
estimate the specific intensity and not the images and/or
SEDs. The resulting noise in the observables is much lower as it only
reflects the noise in the mean specific intensity and no longer the noise
associated to the production of the observables themselves.

This method combining Monte Carlo and ray-tracing can be extended to
any wavelength if the angular 
dependence of the scattering component of the source function is preserved
in the calculations. The Monte Carlo method produces all the
information needed to perform such calculations, as it can give an
estimate of the specific intensity, with its complete angular dependence,
and not only of the mean specific intensity.
However, storing the full spatial, angular and wavelength dependence
of the radiation field requires large amounts of memory which is
currently beyond computational capacities.   

This difficulty can be overcome by invoking successive monochromatic
Monte Carlo runs, which removes the need to store the wavelength
dependence of the specific intensity.
 An initial multi-wavelength Monte
Carlo run calculates the temperature structure in the disc. The SED is
then constructed wavelength by wavelength with 
successive monochromatic Monte Carlo runs that estimate the
specific intensity at each point of the model.

In \mcfost, the specific intensity is then saved for a set of angular
directions (method~1). At the
end of each Monte Carlo run, the scattering emissivity in
any direction is calculated from 
the specific intensity and resolved maps and/or integrated fluxes for
any inclinations are finally obtained by ray-tracing.
This step (monochromatic Monte Carlo run +
ray-tracing run) is repeated over all wavelengths,
without storing the specific intensity at the previous wavelength.

A slightly different method is adopted in \mcmax, where the scattering
emissivity in a given set of directions is stored instead of the
specific intensity itself (method~2).
 The scattering emissivity is the product of the specific intensity by the
scattering phase function, \emph{i.e.} the last term in
equation~\ref{eq:transfert_radiatif}. Each time a packet crosses a cell, its
contributions to the scattering emissivity in the chosen directions
are calculated by multiplying
the packet energy by the local phase function. At the end of each
monochromatic Monte Carlo
run, maps and/or fluxes at the chosen inclinations are produced by
integrating the source function via a ray-tracing method.

Method~1 avoids the expensive calculations of the scattering
emissivity each time a packet crosses a cell but requires a larger
amount of memory to store the angular dependence of the specific
intensity. If the radiation field is stored for
a few specific wavelengths,
it also allows to produce scattered light images and
emission maps at any inclinations, by only running additional ray-tracing calculations.
However, with this method, the angular sampling of the radiation field
must be performed with care, especially when scattering is very
anisotropic. This issue is not encountered with
method~2, which has the same, almost perfect\footnote{only limited by
  the numerical precision}, angular sampling of the radiation
field as classical Monte Carlo methods.

\subsection{Codes description}

\subsubsection{\mcfost}
\mcfost\ is a 3D continuum and line radiative transfer code based on
the Monte Carlo method \citep{Pinte06}. Temperature structures are calculated
using the immediate re-emission concept 
of \cite{Bjorkman01} but with a continuous deposition of energy to
estimate the mean intensity \citep{Lucy99}. 
The code uses a spherical or cylindrical grid, with an adaptive mesh
refinement at the inner edge (based on the opacity gradient) so as to
properly sample the inner radius of the disc.

Several improvements have been implemented on top of the original algorithm
presented in \cite{Pinte06}. In very optically thick parts
of the model, the temperature structure is calculated with a diffusion
approximation method, using the Monte Carlo calculations as limit
conditions. 
Equation~\ref{eq:diff_approx} is solved as an asymptotic limit of the
time dependent diffusion equation, via an implicit scheme to ensure
stability and accuracy.
 The transition between the Monte Carlo and diffusion
approximation regions is set to an optical depth of 1\,000 at the
wavelength where the stellar emission peaks. To avoid
  calculating the propagation of photon packets inside the diffusion
  approximation domain, packets are mirrored at the boundaries of
the Monte Carlo domain.

The temperature structure and radiation field estimated by the Monte Carlo runs
are used to produce images, polarisation maps, and SEDs with a
ray-tracing method, where the emerging flux is obtained by calculating the formal
solution of the radiative equation along rays.

\subsubsection{\mcmax}

The Monte Carlo radiative transfer code \mcmax\ is based on the scheme
of immediate re-emission as proposed by \citet{Bjorkman01}. For the
temperature structure the method of continuous absorption by
\citet{Lucy99} is implemented. The photons are traced in 3D on a
spherical coordinate grid while the geometry of the system is set to
be cylindrically symmetric. For the optically thick regions a modified
random walk procedure is applied in order to make multiple interaction
steps in a single computation \citep[see][]{Fleck84, Min09}.
This has the advantage that the computational speed
is increased significantly while the temperature structure is still
computed with high accuracy. After the Monte Carlo procedure a partial
diffusion approximation is used for these regions in the disc that
received too few photons to determine a reliable temperature structure
\citep[see][]{Min09}. All the observables are
constructed by integrating the formal solution using ray-tracing. In
this way noise on the observables is reduced significantly.  

The spatial grid at the inner edge of the disc is set in such a way
that the optical depth for both the local radiation field and the
stellar radiation are sampled logarithmically.  

\subsubsection{\pinball}
\pinball\ is a Monte-Carlo code that calculates scattered-light images;
it does not calculate the equilibrium temperature or include thermal
re-emission. An earlier version and a pair of simple test cases were
described by \cite{Watson01}. The current version includes
polarisation.  

\subsubsection{\torus}
\torus\ is a 3D continuum and line radiative transfer code based on
the Monte Carlo method \citep{Harries00,Harries04,Kurosawa04}.
Radiative equilibrium is computed 
using the continuous absorption algorithm from \cite{Lucy99}. 
Calculations are performed on a 2D, cylindrical 
adaptive-mesh grid. Storing the opacity information on
an adaptive mesh has particular advantages for the  
problem considered here, since it allows an adequate sampling of the
inner edge of the disc, where the opacity gradient is very steep.
The temperature structure in the central regions of the disc is
computed with a diffusion approximation method.

For scattered light images, the enforced scattering concept
\citep{Cashwell59} as well as the peel-off technique \citep{Yusef84} 
are implemented to reduce the variance.

\paragraph{}
The descriptions of the codes \ProDiMo, \radmc\ and \radical, that have
performed the test cases with isotropic scattering and without
scattering, are presented in section~\ref{sec:grid_codes}.


\section{Benchmark problem}

All the codes in this paper that calculate the thermal equilibrium 
have successfully reproduced
the \P04\ benchmark, from optically thin configurations to optical
depths of 100 in the optical. The test cases presented here are
complementary to those in \P04\ and
are restricted to optical depths higher than 1\,000.
The full description of the benchmark problem, including tabulated
values for the disc density and the dust
properties, as well as the results for all codes  are presented on the
webpage \url{http://www.astro.ex.ac.uk/people/cpinte/benchmark/}. This should
allow additional codes to compare their  results with the ones
presented in this paper.

\subsection{System geometry}
The model consists of a dusty disc surrounding a central star and located
at a distance of 140\,pc.

We consider an axisymmetric flared density structure with a Gaussian vertical
profile $\rho(r,z) = \rho_0(r)\,\exp(-z^2/2\,h(r)^2)$. We use
power-law distributions for the surface density $\Sigma(r) =
\Sigma_0\,(r/r_0)^{-1.5}$  and the scale height $ h(r) = h_0\,
(r/r_0)^{1.125}$ where $r$ is the radial coordinate in the equatorial
plane and $h_0 = 10$\,AU is the scale height at the radius $r_0 =100$\,AU.
The disc extends from an inner cylindrical radius  $r_\mathrm{in} =0.1$\,AU  to an
outer limit $r_\mathrm{out} = 400$\,AU.
The edges of the disc are assumed to be sharp, \emph{i.e.} vertical:
there is nothing inside $r_\mathrm{in}$ and outside $r_\mathrm{out}$
and the density is defined by the previously mentioned power-laws between them.
The dust disc mass is the only parameter varied and takes 4 different values: $3\times10^{-8}$,
$3\times10^{-7}$, $3\times10^{-6}$ and $3\times10^{-5}$\,M$_\odot$.

This configuration was chosen because it represents a more difficult problem to solve than the
test case presented by \P04: the disc extends much closer to the star,
0.1\,AU instead of 1\,AU and the radial gradient of density is much
steeper with a slope of surface density of $-1.5$ instead of 0.125,
leading to much higher densities close to the inner edge of the disc,
and hence much higher disc optical depths.

The star is defined as a uniformly radiating blackbody sphere at a
temperature of 4\,000\,K and with a radius of 2 solar radii.

\subsection{Dust properties}

Dust grains are defined as homogeneous and spherical 
particles with a single size of 1\,$\mu$m and are composed of astronomical
silicates \citep{Weingartner01}. The grain mass density is fixed to 3.5\,g/cm$^{3}$.

The dust optical properties: extinction and scattering opacities (Fig~\ref{fig:opacite}), scattering
phase functions, and Mueller matrices are calculated using the Mie
theory. The resulting midplane optical depth in $I$ band (0.81\,$\mu$m), from the star to the
observer, is ranging from
$1.22\times10^3$ to $1.22\times10^6$ when the disc mass varies from
$3\times10^{-8}$ to $3\times10^{-5}$\,M$_\odot$. For simplicity, in the
following, we will label
the different models $\tau = 10^3$, $\tau = 10^4$, $\tau = 10^5$ and $\tau = 10^6$.

\begin{figure}
  \includegraphics[width=\hsize]{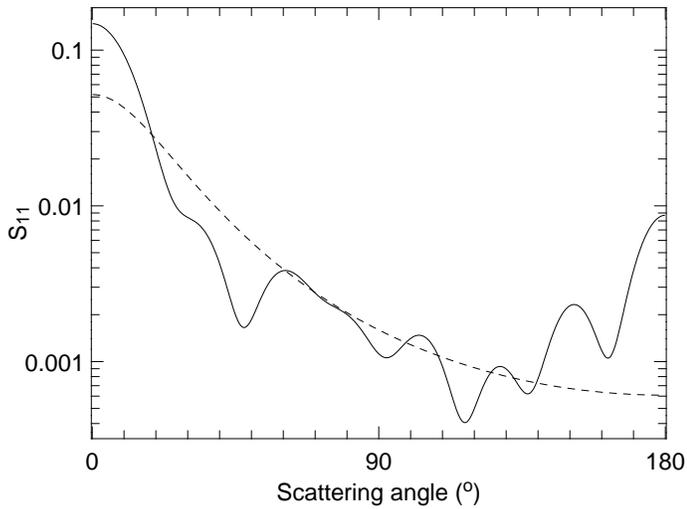}  
  \caption{Mie scattering phase function (first element of the
    Mueller matrix, full line) for the 1\,$\mu$m silicate
    grains at a wavelength of 1\,$\mu$m. The dashed line represents the
    Henyey-Greenstein phase function for the same asymmetry parameter
    $g = \left<\cos \theta \right> = 0.63$. 
\label{fig:phase_fct}}
\end{figure}

In the simplifying case of Mie
scattering, the matrix becomes block-diagonal with only 4 non-zero elements:
\be
\label{eq:mueller}
\left( \begin{array}{r}
I \\
Q \\
U \\
V \\
\end{array} \right)_\textrm{scatt}
=
\left( \begin{array}{rrrr}
S_{11} & S_{12} & 0 & 0 \\
S_{12} & S_{11} & 0 & 0 \\
0    &  0  & S_{33} & S_{34} \\
0 & 0 & -S_{34} & S_{33} \\
\end{array} \right)
\left( \begin{array}{r}
I \\
Q \\
U \\
V \\
\end{array} \right)_\textrm{incident}
\ee
where the individual elements $S_{ij}$ only depend on the scattering
angle and not on the azimuthal angle. The first element
$S_{11}$, also known as the ``phase function'' is plotted in
Fig~\ref{fig:phase_fct} for the wavelength of 1\,$\mu$m used to
calculate scattered light images and polarisation maps.

A spherical grain of 1\,$\mu$m at a wavelength of 1\,$\mu$m is in the middle of
a resonance region, where constructive and destructive interference
within the dust grain results in phase (and polarisation) functions
with strong oscillations. These effects are not observed 
in the case of a grain size distribution (where the oscillations
corresponding to different grain sizes are averaged) or in the case of
more naturally shaped particles, like aggregates. 
However, we choose these dust properties as they represent
a better test case for the radiative transfer codes. The oscillations
in the elements of the Mueller matrix must be seen in the synthetic
maps allowing a more detailed comparison between codes. For comparison
an Henyey-Greenstein phase function with the same asymmetry parameter
is over-plotted in Fig~\ref{fig:phase_fct}.

All the calculations presented in section \ref{sec:results} are done
assuming anisotropic scattering and use the previously presented Mueller matrix.
To compare the results of the Monte Carlo codes with those obtained
from  discrete ordinate codes, the temperature structures and SEDs are also
calculated in the isotropic case (\emph{i.e.} $S_{11}$ is constant),
using the opacities calculated with the Mie theory (section \ref{sec:grid-based}).

\subsection{Maps and SEDs}

SEDs, images and polarisation maps are calculated at 10 inclinations
equally spaced in cosine, \emph{i.e.} for $\cos(i) =$ 0.05, 0.15,
\ldots, 0.85 and 0.95. This corresponds to inclination angles ranging from
18.2 to 87.1$^\circ$ from pole-on.
Scattered light images and polarisation maps are calculated at 1\,$\mu$m.
The pixel scale is 25.61\,mas.pixel$^{-1}$ (\emph{i.e.} 251 pixels for a
physical size of 900\,AU at a distance of 140\,pc). This is roughly a
factor 2 smaller than the pixel scale of the WFPC and ACS cameras
on-board the \emph{Hubble Space Telescope}.

\section{Results}
\label{sec:results}

\subsection{Temperature structures}
\label{seq:temperature}

\begin{figure*}
  \includegraphics[angle=270,width=\hsize]{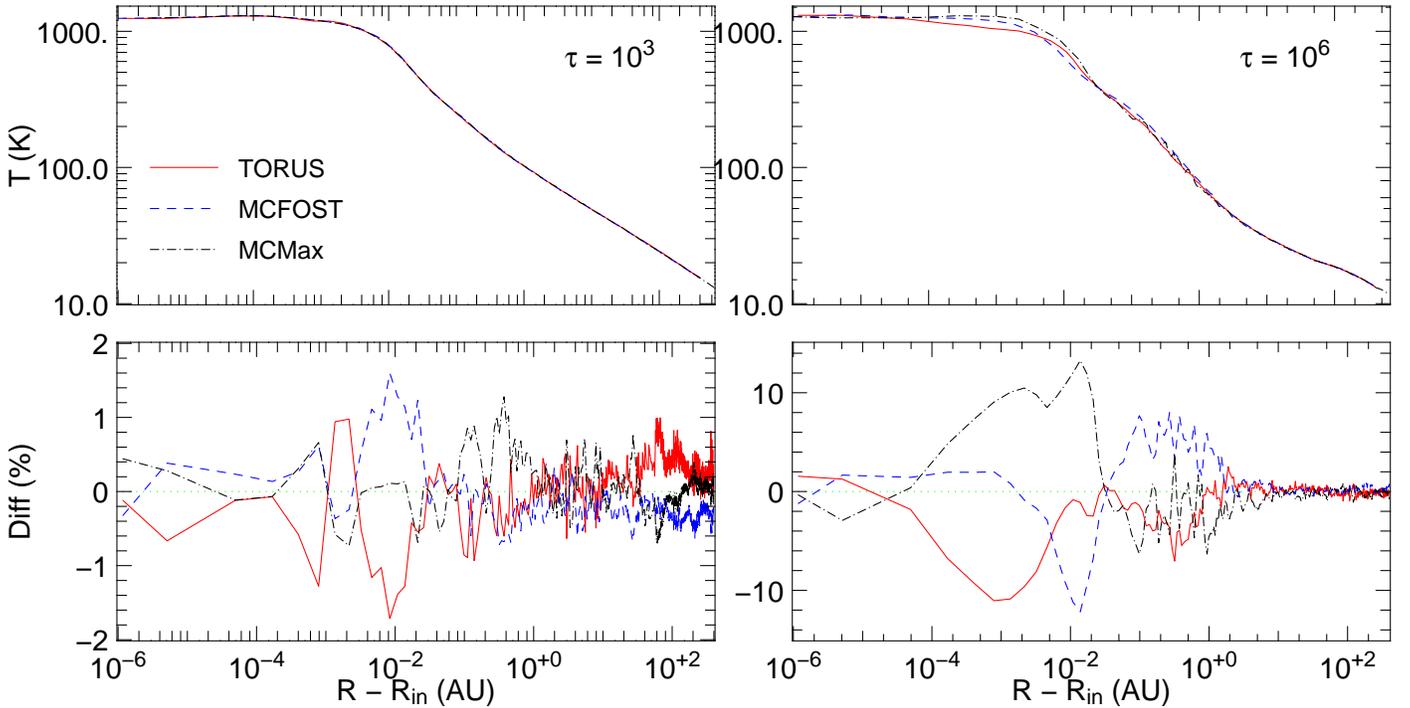}
  \caption{Radial temperature profile in the disc midplane. The left panel corresponds to the $\tau=10^3$ case and the
    right panel to the $\tau=10^6$ case.  Temperature profiles are plotted as
    a function of the distance from the disc's inner radius (and not
    from the star), so as to
    enhance differences between codes. Differences are relative to the average result of the 3
    codes.  The red full lines represent the results of \torus, the
    blue dashed lines, the results of \mcfost\ and the black dot-dashed
    lines the results of \mcmax. The results of \ProDiMo, \radmc\ and
    \radical\ are presented in Fig~\ref{fig:radial_T_iso} (isotropic
    scattering) and Fig~\ref{fig:radial_T_no_scatt} (no scattering).
    \label{fig:radial_T}
  }
\end{figure*}

\begin{figure*}
  \includegraphics[angle=270,width=\hsize]{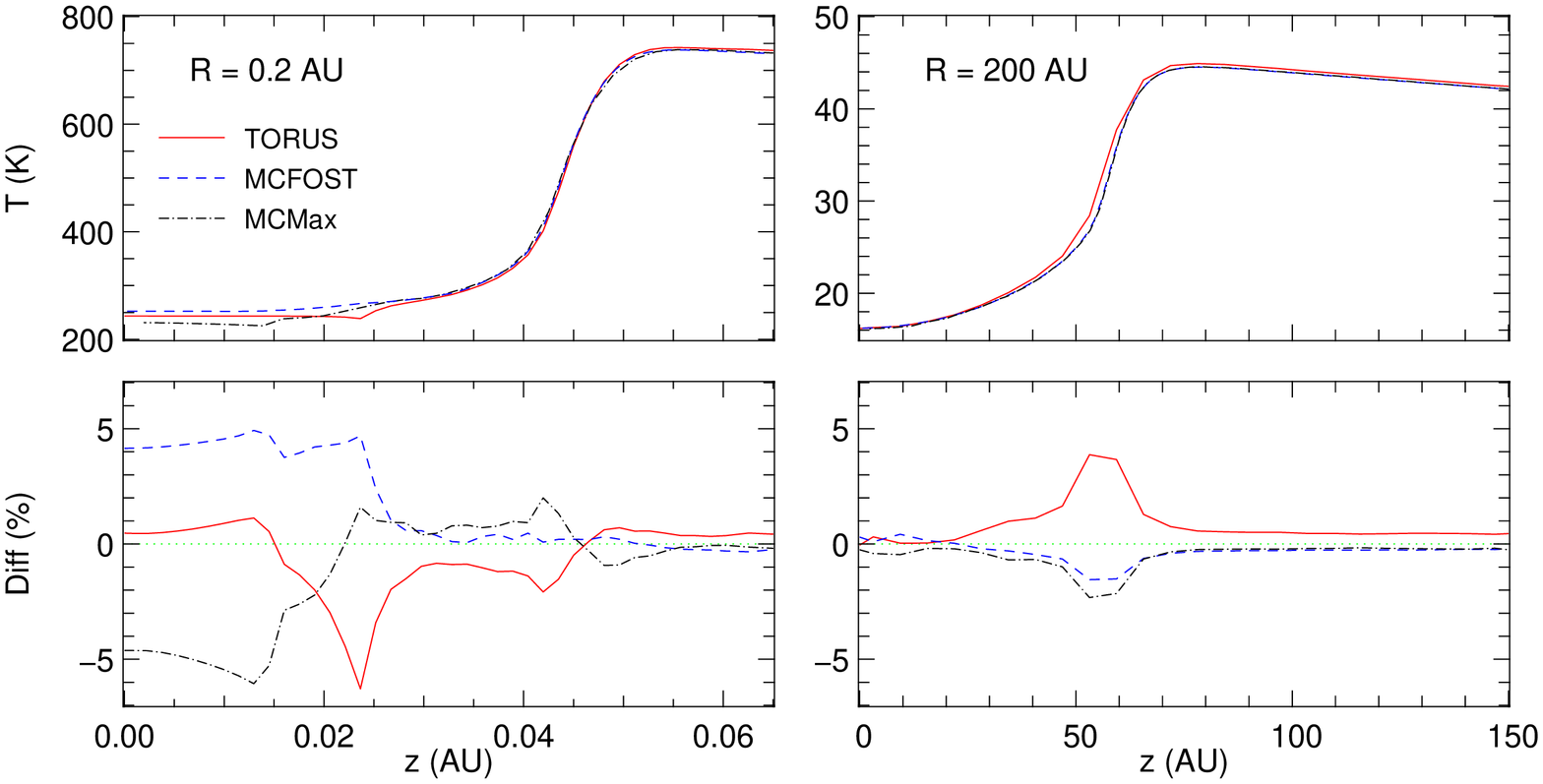}
  \caption{Vertical temperature profiles for the $\tau=10^6$ case. The
    left panel corresponds to a radius of 0.2\,AU and the right panel
    to a radius of 200\,AU. The red full lines represent the results of \torus, the
    blue dashed lines, the results of \mcfost\ and the black dot-dashed
    lines the results of \mcmax. \label{fig:vertical_T}}
\end{figure*}

Figures \ref{fig:radial_T} and \ref{fig:vertical_T} show the
temperature distributions calculated by the different codes, as well
as the difference between codes. Overall, the agreement is very good
with difference almost always smaller than 10\,\%. 

Figure \ref{fig:radial_T}  presents the temperature along the disc
midplane for the $\tau = 10^3$ and $\tau = 10^6$ cases.  
Very close to the 
inner edge, where the disc is directly heated by the star, the agreement between codes is
excellent with maximum differences of the order of 1\,\%.
At large radii ($> 100$\,AU), the disc becomes optically thin at optical
wavelengths in the vertical direction, and the midplane is heated by
the stellar light that is scattered in the surface layers of the
discs. In these regions, the peak-to-peak differences between codes
remain below 1.5 and 
3\,\% for $\tau = 10^3$ and $\tau = 10^6$ cases, respectively. This shows that
all codes deal smilarly with the redistribution of energy by
anisotropic scattering. This is confirmed by the vertical cut at a
radius of 200\,AU (Fig~\ref{fig:vertical_T}, right panel), where
the differences are of the order of 1\,\%, except at the turnover
point from optically thin to optically thick (the place where the
temperature suddenly drops) where differences reach 5\,\%.

Differences in the radial temperature profile are larger between the inner
edge and 1\,AU, \emph{i.e.} in regions
were the stellar radiation does not penetrate, even via
scattering. In these regions, the heating mechanism is the dust
re-emission by the upper layers of the disc, which represent
the most difficult case for RT codes.  
Nevertheless, the peak-to-peak differences between codes remain limited, smaller
than 4\,\% for the $\tau = 10^3$ case.
For the $\tau = 10^6$ cases, differences are most of the time smaller
than 10\,\%, except in a very small regions between $r_\mathrm{in} +
10^{-4}$\,AU and $r_\mathrm{in} + 10^{-2}$\,AU where the maximum
difference is 20\,\%. Differences remain also very small
in the vertical direction as shown in the left panel of
Fig~\ref{fig:vertical_T}. They are below 5\,\% from the midplane
 up to the disc surface, where they become smaller than 1\,\%.

\subsection{SEDs}

\begin{figure*}
  \includegraphics[angle=270,width=\hsize]{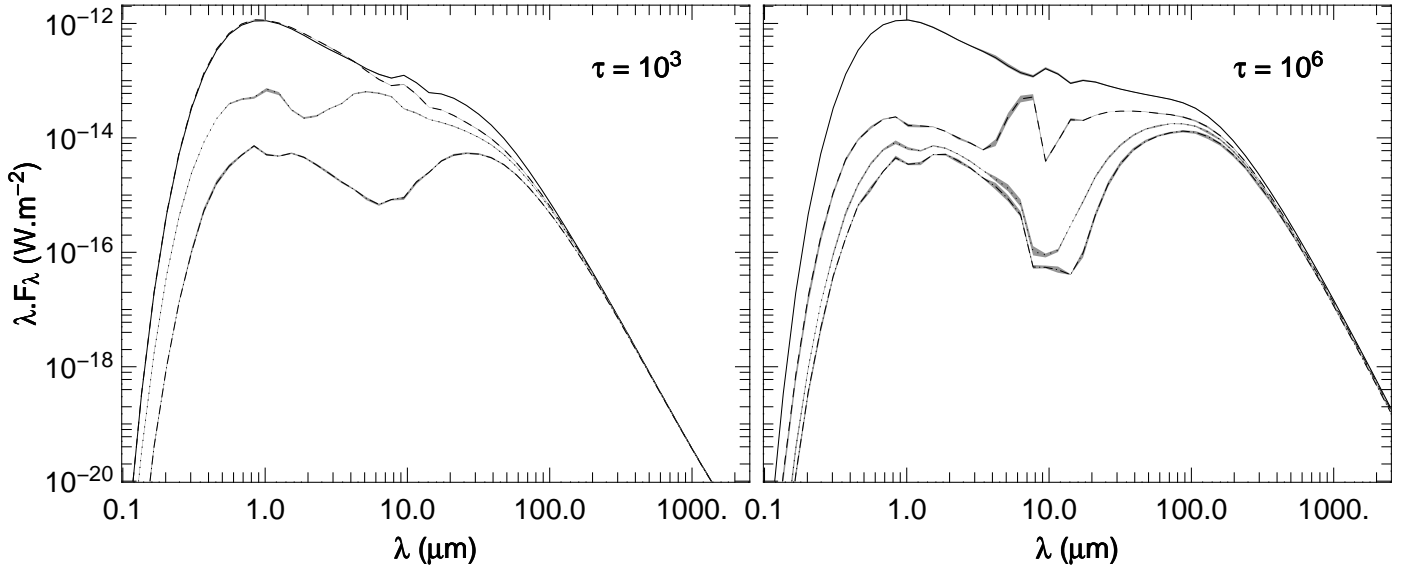}
  \caption{Spectral energy distributions for i=18.2, 75.5, 81.4 and
    87.1$^\circ$ (full, dashed, dotted and dot-dashed lines
    respectively). The left panel corresponds to the $\tau=10^3$ case
    and the right panel to the $\tau=10^6$ case. The lines show the
    average of the results of \mcfost, \mcmax\ and \torus\ and  the grey
    envelope around each line represents the
    range of results obtained by the different codes.\label{fig:SED}}
\end{figure*}

The emerging spectral energy distributions for the $\tau=10^3$ and $\tau=10^6$
models are presented in Fig~\ref{fig:SED} for different inclinations
ranging from an almost face-on ($i=18.2^\circ$) to an almost edge-on disc
($i=87.1^\circ$). 

The shape of the SED is strongly dependent on the inclination, 
moving from a stellar photosphere plus disc excess for low
inclinations to a double-bumped SED, typical of very close to edge-on
systems, when the stellar photosphere is obscured by the disc. 
The disc being optically thick at short
wavelengths, the visible and near-IR stellar light is blocked and the
emission in this wavelength range 
is dominated by the stellar scattered light coming from the disc. At longer
wavelengths ($>10-12\,\mu$m)  the dust emission dominates, resulting
in a steep positive slope and a double-bumped SED.
Not surprisingly, the transition between these two
characteristic SEDs also depends on the disc opacity. For instance, for
an inclination of 75.5$^\circ$, the star is still seen directly in the
$\tau=10^3$ case, whereas it is already strongly obscured in the
$\tau=10^6$ case. 

It is interesting to note that for the most edge-on case,
the flux in the optical is almost independent of the disc opacity. 
Similarly, in the most
optically thick case, there is only small differences in the SEDs at
81.4 and 87.1$^\circ$. Indeed, when the
star is completely obscured, the flux is dominated by 
stellar light that has scattered on the upper layers of the outer
disc. This scattered light is mainly a
function of the dust properties and of the scattering geometry but not
of the optical depth in the line of sight. In these cases, optical and near-infrared
photometry cannot be used to get estimates of the extinction by
dust of the central object. 

At low inclinations, the characteristic 9.8$\,\mu$m
amorphous silicate feature is seen in emission. At higher inclinations, for
example $i=75.5^\circ$ for the $\tau=10^6$ case, the feature is now 
observed in absorption over the continuum emission of the disc.
It should be noted that, at very high inclinations, although the 
dip is roughly centred on the silicate feature, it is not
associated to it as demonstrated by the exceptional 
breadth of the feature. The disc is now
optically thick in the silicate feature but also in the adjacent continuum,
and the absorption feature from the silicates vanishes.

At long wavelengths ($> 500\,\mu$m), the
disc become optically thin in most of its parts and the emerging flux
 no longer depends on the system inclination.

\begin{figure*}
  \includegraphics[angle=270,width=\hsize]{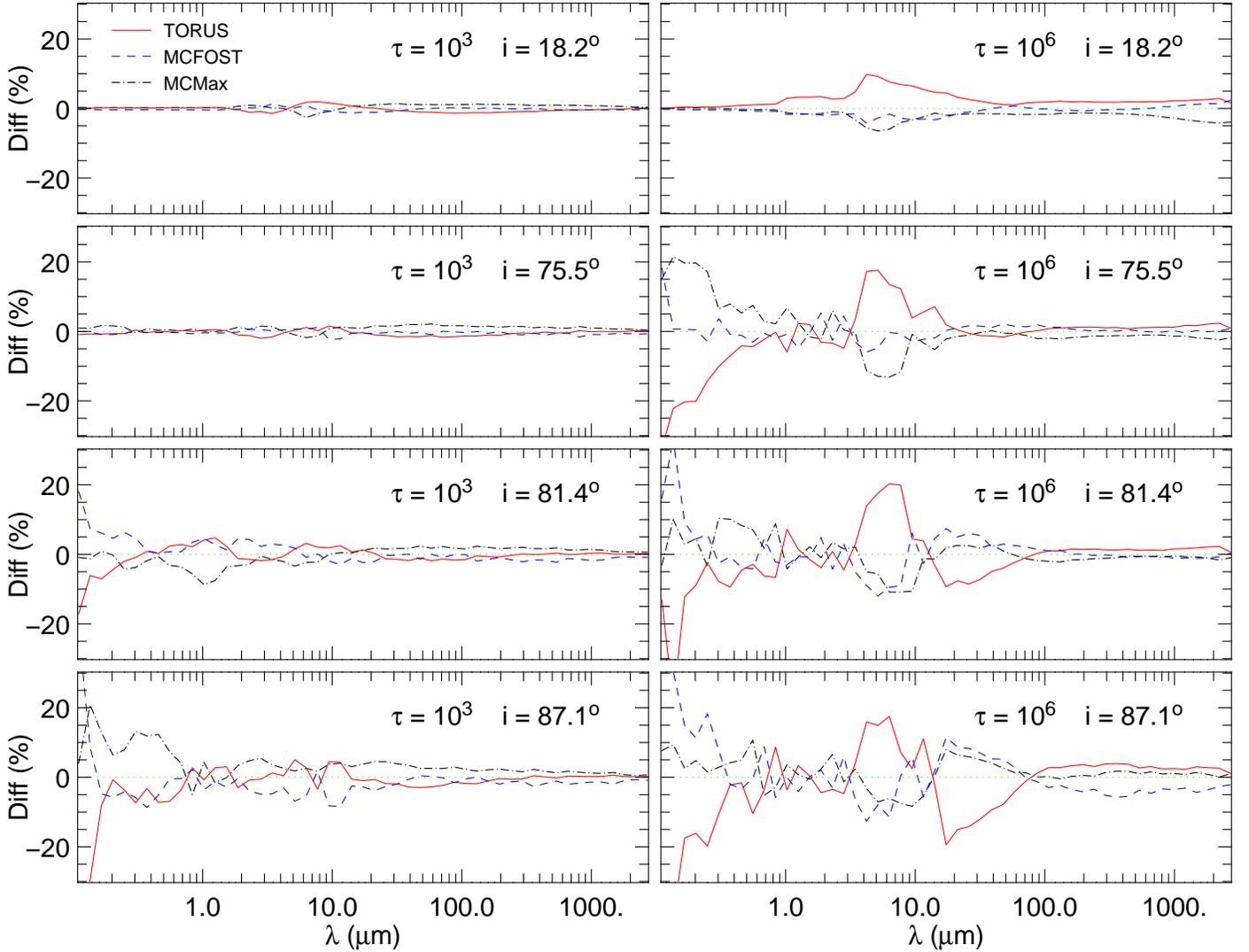}
  \caption{Differences in the SEDs obtained by the different
    codes. The disc opacity and inclination are given on top of each
    panel. The average result of the codes is taken as a
    reference. The red full lines represent the results of \torus, the
    blue dashed line, the results of \mcfost\ and the black dot-dashed
    line the results of \mcmax.  \label{fig:diff_SED}}
\end{figure*}

Overall, the differences between codes are small in all cases, which is illustrated
by the thinness of the grey envelopes around the lines in
  Fig~\ref{fig:SED}, representing the range of the results obtained
  by the different codes.
Fig~\ref{fig:diff_SED} gives a more detailed view of the differences between the codes for
the various inclinations and optical depths.

The left column show the results for the results for the $\tau=10^3$
case. When the star is seen directly ($i=18.5$ and $75.5^\circ$),
the agreement between the three codes is excellent, with peak to peak difference
smaller than 5\,\% over the whole wavelength range.
Closer to edge-on, differences remain smaller than 10\,\%, except at
very short wavelengths ($< 1\,\mu$m) where the contribution from
scattered light is dominating the SED. As the wavelength becomes
shorter, the scattering becomes more forward
throwing and calculations are very sensitive to the
angular sampling of the scattering phase function of the codes.
  As a result, the agreement between codes
becomes worse at short wavelengths. Differences remain smaller than
15\,\% down to $0.2\,\mu$m and significant differences are only
observed around $0.1\,\mu$m.

For the $i=81.4^\circ$ case, the disc is seen at a grazing incidence and
the optical depth from the star to the observer is varying strongly
across the stellar disc, from $\approx 2$ at the top of the stellar
surface to $\approx 200$ at the bottom. 
In this case, using a point source for the star does not
  provide the correct result, and special care
must be taken to resolve the stellar photosphere.
In the case of a uniformly radiating sphere, as presented in this benchmark, the origin of the 
photon packets is uniformly distributed on the stellar surface, and
a uniform distribution in the cosine of the
angle between the photon direction and the normal to the surface at the
point of origin is used to set the initial propagation direction of packets.

For the $\tau=10^6$ case (right panel of Fig~\ref{fig:diff_SED}), the
results are similar to the ones for the lower optical depth case, but
with larger differences in the near- and mid-infrared. This part of
the SED is one of the most challenging 
wavelength range for RT codes, where different contributions (thermal
emission from the inner disc seen directly or through the outer disc,
direct or scattered stellar light, scattered thermal emission from the
disc) can dominate the emerging flux depending on the system geometry and dust
opacities. Furthermore, most of the flux in this wavelength regime is coming
from the inner edge of the disc and the output spectrum is
very sensitive to the grid resolution adopted by the different codes.
\mcmax\ and \mcfost\ present very good agreement over all the
wavelength range, including in the near- and mid-infrared regime, and for all
inclinations. \torus\ shows slightly larger differences, probably due
to a lower spatial resolution at the inner edge. We note that the
spatial resolution of the \torus\ code is limited by the maximum cell
depth in the AMR grid, which is currently set to 30, correponding to a
dynamical range of $2^{30}$ (a limit dictated
by numerical precision in the photon path integrator).  
These differences are maximum around $10\,\mu$m and vary from 15\,\% in the
low inclination case up to 30\,\% when the inclination is increasing.

\subsection{Scattered light images and polarisation maps}

\begin{figure*}
  \includegraphics[angle=270,width=\hsize]{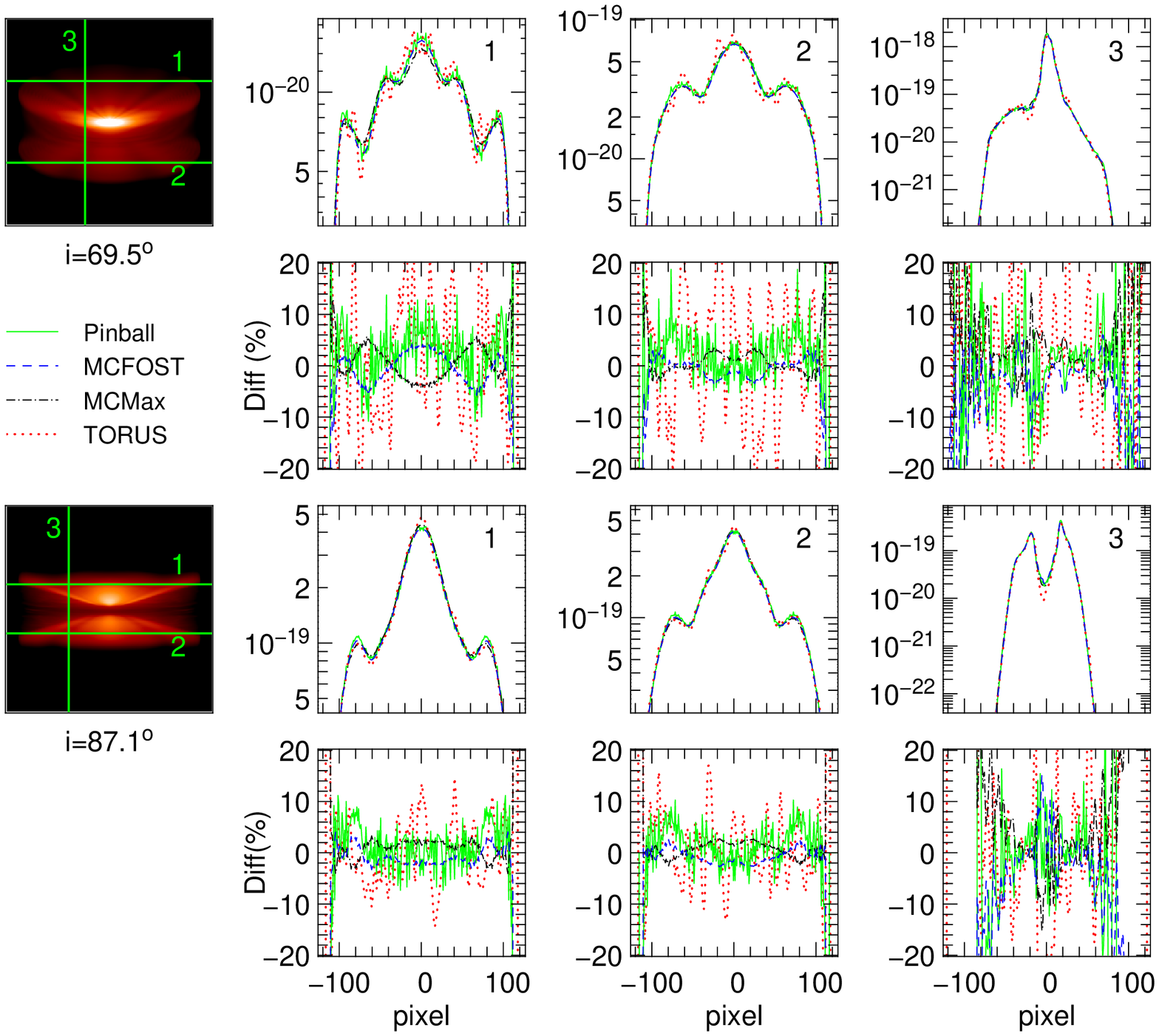}
  \caption{Scattered light images for two inclinations in the $\tau =
    10^6$ case. The green full line presents the results of \pinball,
    the blue dashed line
    the results of \mcfost\, the  black dot-dashed line the results of \mcmax\ and
    the red dotted line the results of \torus. The 3 panels
    on the right show brightness profiles along the cuts plotted in
    the left panel. The cuts are 11 pixels large. Differences are
    plotted relative to the average of results of
    \mcfost\ and \mcmax, which present much lower noise. \label{fig:images}}
\end{figure*}

Fig~\ref{fig:images} presents the scattered light images of the most
optically thick disc for $i=69.5^\circ$ and $i=87.1^\circ$. The
synthetic maps clearly display the effects of the anisotropy of the
scattering. The oscillations in the phase function
(Fig~\ref{fig:phase_fct}) are directly observed in the maps. 

The three panels on the right present the flux obtained by the various
codes, and the corresponding differences, along horizontal and
vertical cuts in the images. The codes agree to within 10\,\% where the
flux is significant. \torus\ shows slightly larger deviations that are
due to a larger Monte carlo noise in the images. 
All codes predict the same
oscillations as a function of the position, indicating that the
implementation of the scattering phase function is correct in all
codes. At greater radii, larger
differences are observed due to the various grid geometries used by
the different codes. For instance,
some codes use a spherical grid whereas other codes use a
cylindrical grid with a vertical cut-off in density. The sampling
of the density structure at the outer edge of the model domain is then
slightly different for each code. This results in systematic
differences between codes, but only in the regions of the synthetic maps where the flux is extremely low.

The vertical cut (\#\,3)  samples the disc ``dark lane'', corresponding
to the optically thick midplane.  In this region, the flux is dominated
by photons that have scattered several times in the disc before reaching the
observer.
The  very good agreement between codes in the dark lane indicates that
all of them deal properly with multiple scattering.

\begin{figure*} 
  \includegraphics[angle=270,width=\hsize]{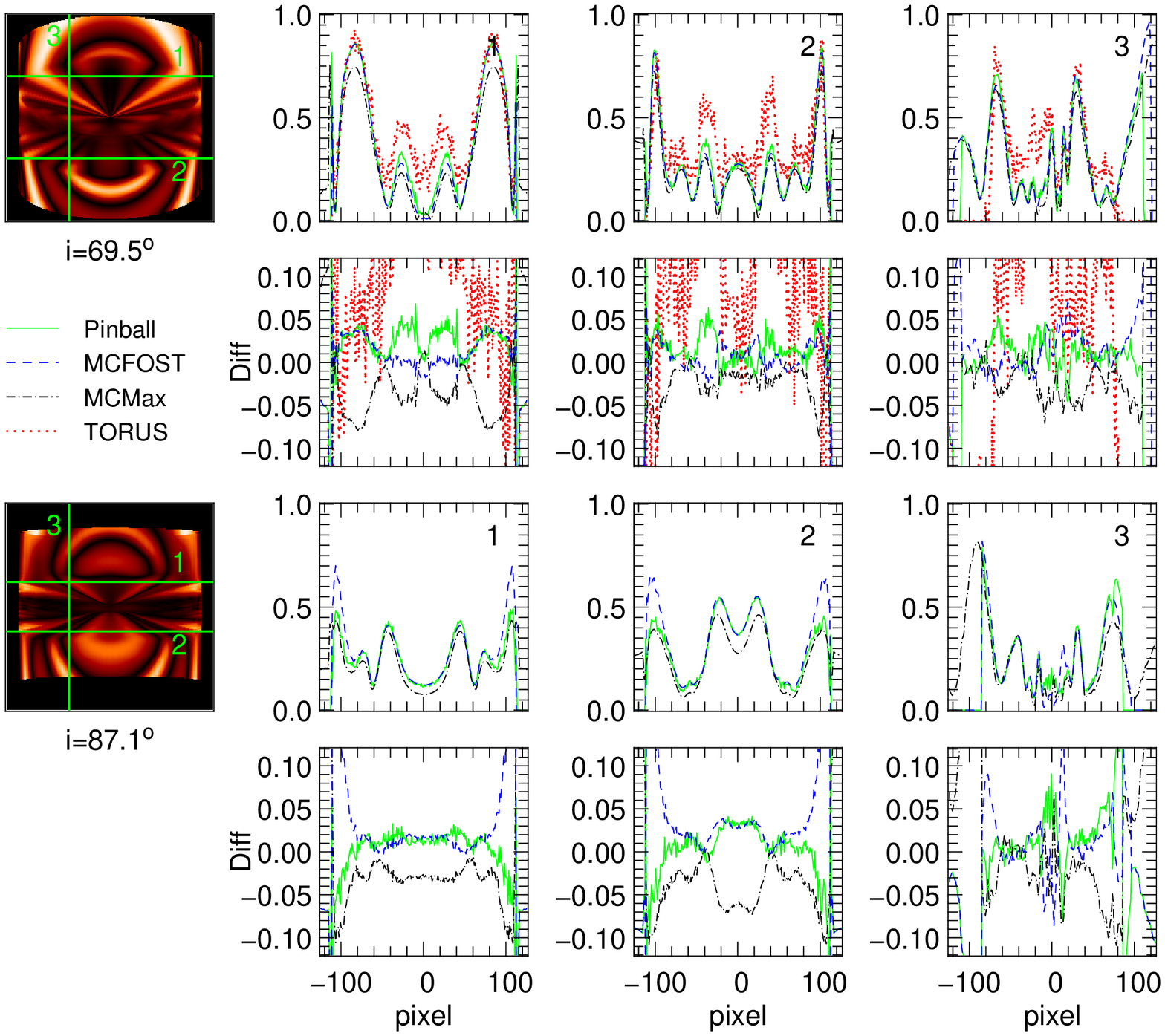}
  \caption{Linear polarisation maps for two inclinations in the $\tau
    = 10^6$ case. The panels and labels are the same as in
    Fig.~\ref{fig:images}. Green full line: \pinball, red dotted line:
    \torus, blue dashed line: \mcfost, black dot-dashed
    line: \mcmax. Cuts are 11 pixels large. The reference is the average
    of  the results from \mcfost, \mcmax\ and \pinball. \torus\ 
    calculations are not presented for the highest inclination calculations due to
    a low signal-to-noise.
    \label{fig:pola}}
\end{figure*}

Fig~\ref{fig:pola} presents the polarisation maps for the same
configurations as in the Fig.~\ref{fig:images}. The maps show complex
patterns due to the strong variations of the elements of the Mueller
matrix with the scattering angle.
All codes produce very similar maps (see horizontal and vertical cuts), at both inclination angles,
with differences smaller than 5 points of polarisation degree in the
central regions of the maps, \emph{i.e.} in regions where the flux is
large enough to allow resolved polarisation measurement in actual
observations. At greater radii, differences become slightly larger,
but they remain limited to 10 points of polarisation degree.
\torus\ shows larger deviations, due to a larger Monte Carlo noise in the
simulations, which biases the polarisation degree towards larger
values. In regions where the polarised flux is large, the
agreement between \torus\ and the other codes is also very good. For the highest inclination, the results from \torus\
  are not shown due to a low signal-to-noise. 

\section{Comparison with discrete ordinate codes}
\label{sec:grid-based}

All results presented thus far were obtained using Monte Carlo codes. To
further compare the predictions of radiative transfer codes, we present,
in this section, results obtained with discrete ordinate codes and discuss
their limitations compared to those of Monte Carlo
codes. Discrete ordinate
codes solve the radiative equation along
predetermined sets of directions. This integration can be performed
with ``long'' or ``short characteristics'' and various schemes can be
used to iterate between the temperature structure and specific
intensity (or its moments), like the Accelerated Lambda Iteration
(ALI) or Variable Eddington Tensor (VET) methods (see
\citealp{Mihalas84}).

The benchmark problem presented in this section is the same as in the previous sections but with
the following modifications:
\begin{itemize}
\item the star is now considered as a point source, but with the same
  spectrum and luminosity as previously defined.
\item the scattering is assumed either to be isotropic (but with the
  same opacities as before), either to be negligible (the scattering
  opacity is set to 0).
\end{itemize}
These tests have been defined to allow a larger number of codes to reproduce the
calculations. They correspond to an extension of the \P04 benchmark to
optical depths higher than 100, but without any further complexity. 
In this paper, they have been calculated by at least one of the
previously tested Monte Carlo code and by two additional
discrete ordinate codes and
one additional Monte Carlo code.  

\begin{figure*}
  \includegraphics[angle=270,width=\hsize]{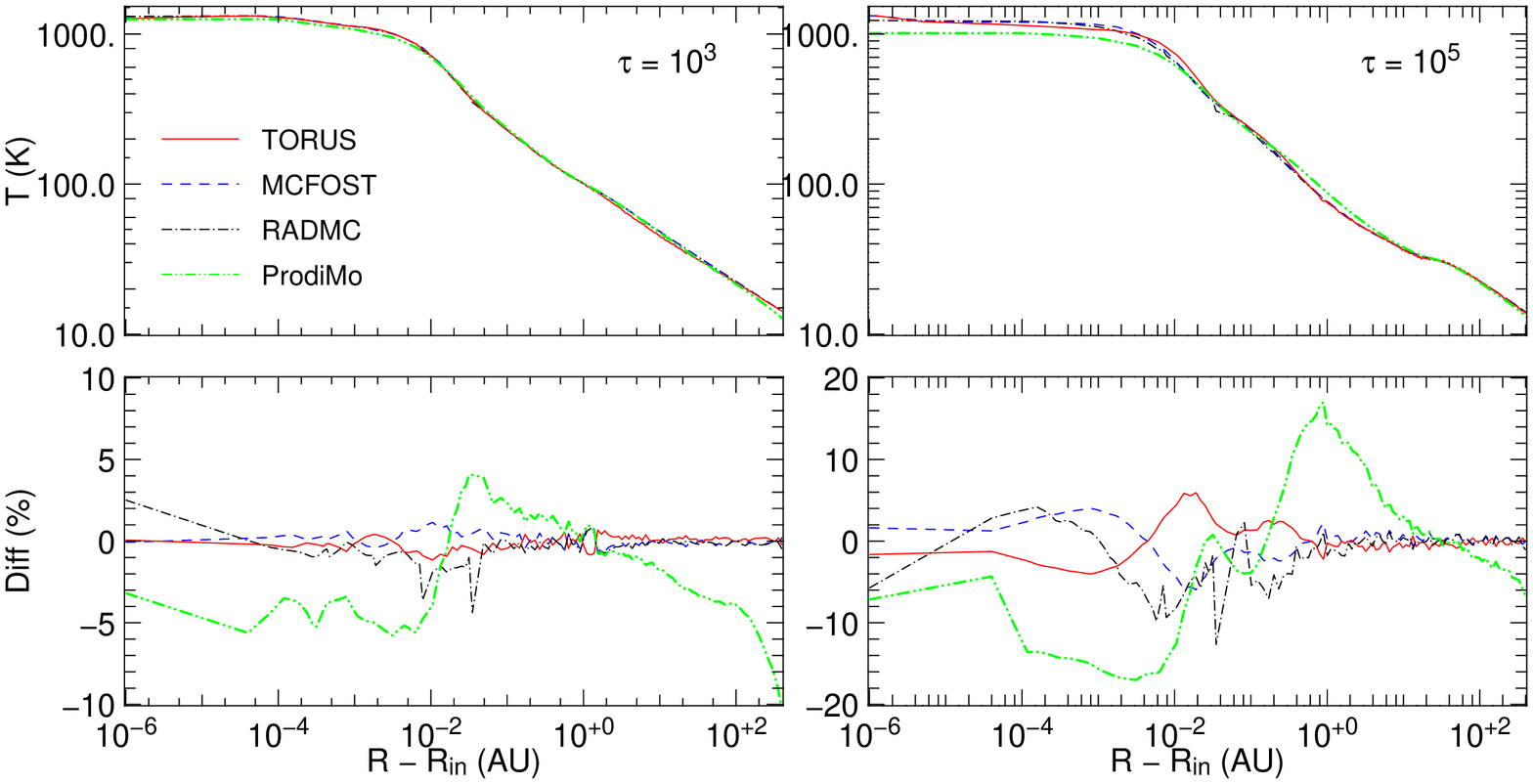}
  \caption{Radial temperature profile in the disc midplane (isotropic
    scattering). The left panel corresponds to the $\tau= 10^3$ case
    and the right panel to the $\tau=10^5$ case. The full red lines
    represent the results of \torus, the blue dashed lines, the
    results of \mcfost, the black dot-dashed lines the results of
    \radmc\ and the green dot-dot-dashed line the results of \ProDiMo.
    Differences are relative to the average results of the Monte Carlo
    codes \mcfost,
    \radmc\ and \torus.  
    \label{fig:radial_T_iso}}
\end{figure*}

\begin{figure*}
  \includegraphics[angle=270,width=\hsize]{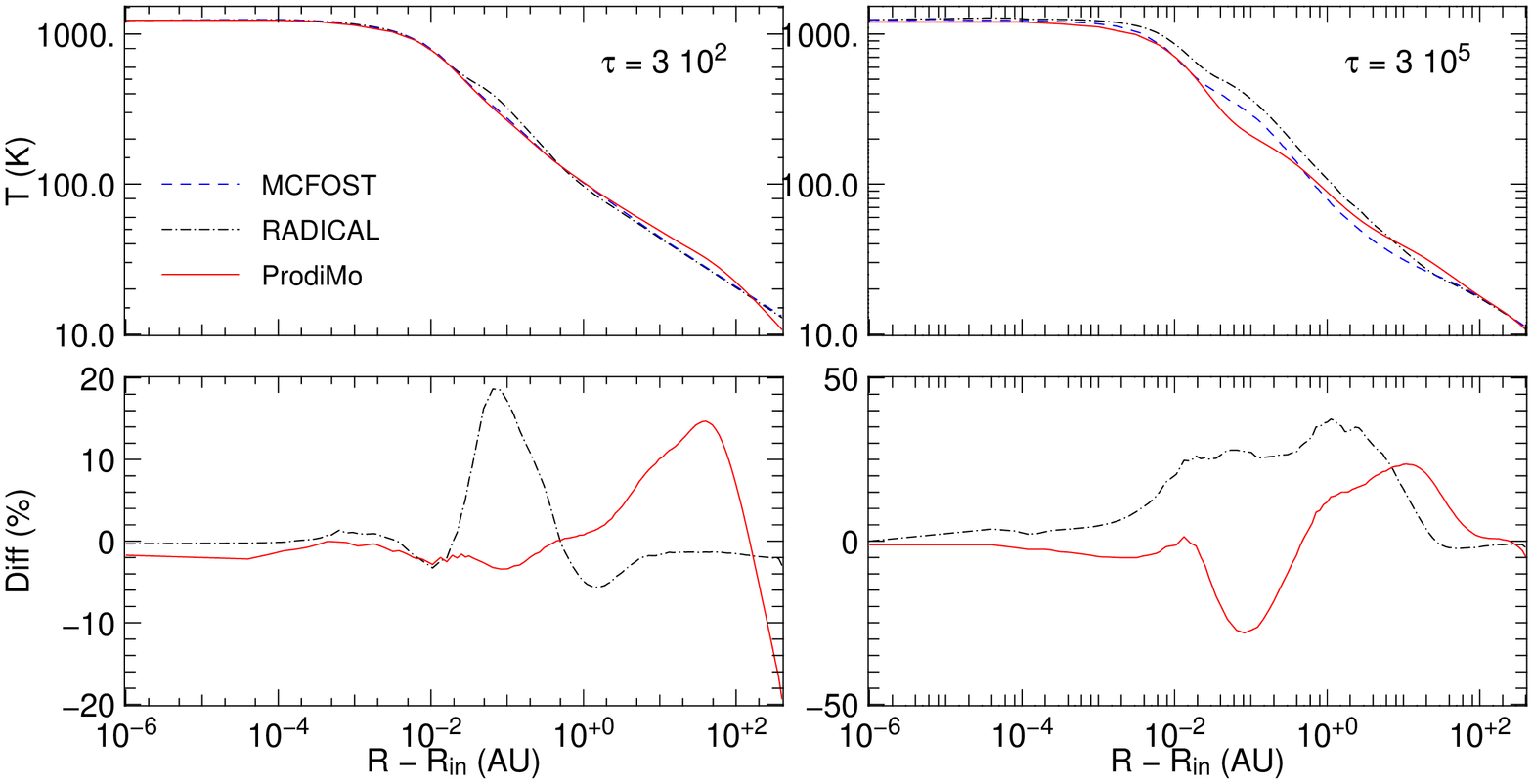}
  \caption{Radial temperature profile in the disc midplane (no
    scattering). The left panel corresponds to the lowest disc mass of
    $3\,10^{-8}$\,M$_\odot$, and the right panel to the highest disc
    mass $3\,10^{-5}$\,M$_\odot$. The corresponding optical depths at
    1\,$\mu$m are lower than in the previous cases because we fixed
    the scattering opacity to zero. The red full lines represent the
    results of \ProDiMo, the blue dashed lines the results of
    \mcfost\ and the black dot-dashed lines the results of \radical.
    The Monte Carlo code \mcfost\ is taken as reference.
    \label{fig:radial_T_no_scatt}
  }
\end{figure*}

\subsection{Code description}
\label{sec:grid_codes}

\subsubsection{\ProDiMo}
\label{sec:prodimo}
\ProDiMo\ is an acronym for \underline{Pro}toplanetary \underline{Di}sk
\underline{Mo}del \citep{Woitke09} which
consistently solves the chemistry, the heating/cooling balance of the
gas, the dust radiative transfer and the vertical stratification of
protoplanetary discs, mainly for the purpose of interpreting far IR to
mm gas emission lines.

For realistic gas models, it is essential to calculate the dust
temperature structure in the disc as well as the transport of UV photons
including scattering, which drive the photo-chemistry. Furthermore,
radiative pumping by continuum radiation changes the non-LTE
population of atoms and molecules and have an important impact on
the cooling rates. These strong physical couplings necessitate to
solve a full 2D dust radiative transfer as one module in a global
iterative procedure.  It is this radiative transfer module inside
\ProDiMo\ that participates in this benchmark test. Its basic task is to
provide $\Td(r,z)$ and $J_\lambda(r,z)$ for the gas modelling -- it is not
meant for the interface to dust observations (SEDs, scattered light
images etc.).

\ProDiMo\ solves the frequency-dependent 2D dust continuum radiative
transfer of irradiated discs by means of a simple, ray-based,
long-characteristic method.
 From each grid point in the disc, a limited number of rays (here
172) are traced backwards, while solving the radiative transfer
equation with isotropic scattering. The setup of the ray directions is
critical for the optically thin parts of the disc, in particular at
near IR wavelengths where the illumination originates from small and
far hot regions. This is done in a manual fashion in \ProDiMo\ to ensure there are
more rays pointing toward the hot inner regions than toward the cooler
interstellar side. One central ray pointing toward the star is
reserved and covers the solid angle occupied by the star.

Instead of a treatment with a large number of wavelength
grid points, \ProDiMo\ uses a coarse wavelength grid
$\{\lambda_k\,|\,k=0,...,K\}$ (here $K\!=\!24$) from $100\,$nm to
$1000\,\mu$m, and treat the opacities, intensities and source
functions with band means, e.g.\ $B_k(T) =
\frac{1}{\Delta\lambda_k}\int_{\lambda_{k-1}}^{\lambda_k} B_\lambda(T)\,d\lambda$ where
$\Delta\lambda_k=\lambda_k-\lambda_{k-1}$. One radiation transfer iteration takes
about 4 seconds for a low resolution $20\times20$ grid, and about 70
seconds for a $70\times70$ grid on a single-processor 2.66\,GHz Linux
machine, which is comparable to the computational efforts taken to
solve the disc chemistry.

In order to solve the condition of radiative equilibrium and the
scattering problem, a simple $\Lambda$-type iteration is applied. The
source functions are pre-calculated on the grid points and fixed
during one iteration. During the ray-tracing, the opacities and source
functions are interpolated from the grid point values.  After having
solved all rays from all points in all frequency bands, the mean
intensities are updated and the dust temperatures are
re-calculated. Without further accelerations, this $\Lambda$ iteration
converges only for problems up to a midplane optical depth of about
100.  In order to accelerate the convergence, we apply the procedure of
\cite{Auer84} known as ``Ng''-iteration. This enables us
to solve radiative transfer problems up to $\tau\!=\!10^3\ldots10^4$,
depending on the geometry of the model.

For higher optical depths, we apply an approximate procedure following
an idea of C.P.~Dullemond which consists in reducing the dust density
in the central 
midplane regions in the following way. For every vertical column
(considering the downward direction) we do not increase the dust
density any further once a certain critical optical depth at $1\mu$m
is reached ($\tau_{\rm crit}\!\approx\!10$), provided that the radial
optical depth is also $>\!\tau_{\rm crit}$. With this ``trick'', we
can manage test problems up to $\tau\!=\!10^5$ with this code.

\subsubsection{\radmc}
\radmc\ is a Monte-Carlo based continuum radiative transfer code for 2-D
axisymmetric configurations such as circumstellar discs and envelopes.  The
basic algorithm is that of \cite{Bjorkman01}, but with a continuous
deposition of energy instead of the discrete deposition as described in the
original paper. In this way the temperature profile is also smooth in the
very optically thin regions of the model. The temperature corrections are
not computed every time a photon package enters a cell and leaves some
energy, but only if the energy deposited since the last temperature update
is larger than some threshold value. In this way the not so cheap
temperature update is done only when needed. Of course, the higher this
threshold, the faster the code is but the less reliable the
result. Typically a threshold of a few percent is taken, meaning that if the
energy deposited in the cell has increased by more than a few percent of the
energy as it was during the last temperature update, then a new temperature
update is done. Once the main Monte Carlo process is over, the spectra and
images from \radmc\ are made using a post-processing step: the ray tracing
program \raytrace\ uses the dust temperature and isotropic scattering source
function computed by \radmc\ to calculate the formal solution of the transfer
equation along rays through the model. This yields images at every discrete
wavelength bin. By integrating over the images one obtains a flux at each
wavelength, \emph{i.e.}\ a spectrum. Care is taken to arrange the pixels of the
images such that all the flux is captured, both from the very outer regions
and from the very inner regions. This is done using ``circular images'',
described in detail in \cite{Dullemond00}. Because the spectra and
images are computed as a post-processing step, as opposed to the more
classical photon collection during the Monte Carlo process itself, it is
hard to include full non-isotropic scattering. To keep the flexibility to
view the object from every angle and at every wavelength without having to
repeat the \radmc\ run, one has to store the entire scattering source function
$S(r,\theta,\mu,\phi,\lambda)$, where $\mu$ and $\phi$ are the local directional
coordinates. Such a 5-dimensional array is extremely large and requires of
the order of a gigabyte of disc space, which is not very practical. Instead
one could prescribe before calling \radmc\ at which angle you wish to view the
object, removing the need to store the source function also as a function of
$\mu$, $\phi$, meaning we have a 3-D array which is much easier to
store. This is done in several other codes in this paper. This is not
implemented in \radmc.

\subsubsection{RADICAL-VET} 
The code \radical\ is a classical discrete ordinate method for radiative
transfer, \emph{i.e.}\ it is not based on a Monte Carlo approach and is therefore
completely deterministic. It is based on methods that are routinely used in
models of stellar atmospheres, with some adaptions. The algorithm is that of
Variable Eddington Tensors (VET), which is a multi-dimensional version of
the method of Variable Eddington Factors (VEF) described in the book by
\cite{Mihalas84}. A 1-D version of this method, with special
application to the kind of continuum radiative transfer problems encountered
in protoplanetary discs, was described in \cite{Dullemond02}.
For such 1-D geometries the method is extremely accurate and
efficient. It works well and converges quickly for optical depths ranging
from small ($\ll$1) to extremely large ($\gg 10^6$). The \radicalvet\ code is
a 2-D version of this algorithm. 
For 2-D or 3-D geometries the method
has some numerical difficulties related to the computation of the flux-mean opacity in
regions of extremely low flux (e.g. the midplane of a passive
irradiated disc). In practice, 
however, these difficulties are not fatal, although they could lower
the reliability of the 
method in such flow-flux regions.
For further details on the VET method the reader is referred to
\cite{Dullemond02}, even though this paper describes only the
1-D version of this method.

\subsection{Results  and discussion}

The midplane temperature profiles are presented in
Figs~\ref{fig:radial_T_iso} and \ref{fig:radial_T_no_scatt} for the
cases with isotropic scattering and no scattering
respectively. Convergence was not properly reached for some codes for
the highest mass case with isotropic scattering and we present the
results for the $\tau=10^5$ case instead.

The agreement between Monte Carlo codes is again very good in the
isotropic case (Fig~\ref{fig:radial_T_iso}), with peak-to-peak
difference smaller than a few
percents in the $\tau=10^3$ case and smaller than 15\,\% in the 
$\tau=10^5$ case. Differences with discrete ordinate codes are significantly largers.

 This paper states an important verification test for the method of
 reducing the dust density implemented in \ProDiMo\ and described in 
section~\ref{sec:prodimo}.  It
 shows that the error of this procedure in comparison to the results of the
latest Monte Carlo codes combined with diffusion solvers is smaller than
about 20\,\% in the
midplane regions.  We note however that the relative precision in
general is much better than in the midplane.  The average
difference\footnote{over 500 representative points in the grid}
between \ProDiMo\ and 
\mcfost\ temperature structures, $|T_\ProDiMo - T_\mcfost| /
T_\mcfost$ is smaller than  5\,\% (1~sigma-deviations) in all cases.

The benchmark test has revealed three major problems quite typical for
ray-based codes. First, the limited number of fixed rays causes some
small artifacts of the temperature structure in the optically thin distant
midplane -- these problems can be solved by using a larger number of
rays which is, however, computationally expensive. Second, the precise
temperature determination in the optically thick core of the disc is
hard with simple discrete ordinate codes like \ProDiMo. The numerical
solution of the radiative transfer equation has always some
discretisation errors superimposed, and in an optically thick
situation it is the small difference $I_\lambda-B_\lambda$ that determines the
next temperature iteration. This eventually limits the quality of the
``forecast'' by the Ng-iteration, and disables the convergence for very
optically thick problems.  Third, the results close to the midplane
suffer from the very large gradients present close to the inner
boundary, and the results depend on the numerical details, e.g.\ how
to interpolate the source function.

Comparisons with the \radical\ code illustrates some of the
difficulties of the VET method at very high optical depths. \radical\ 
overestimates the midplane temperature in the central regions of the
disc by about 20\,\% and 40\,\% for the low and high mass cases
respectively (Fig~\ref{fig:radial_T_no_scatt}). 
The 2-D geometry introduces a number of
difficulties which make the 2-D VET algorithm less stable and less reliable
than its 1-D version. The main problem is the choice of discrete angular
coordinates at each grid point. For the formal transfer \radicalvet\ uses the
method of Short Characteristics. The choice of the angular distribution of
these short characteristics is essential for the reliability of the result.
In \cite{Dullemond00} a good choice was described, but in the end
no choice is perfect and the reliability of the results may depend on this.
Another difficulty is that the VET method uses {\em flux-mean opacities}
which are the generalisations of Rosseland mean opacities. By using
flux-mean opacities instead of Rosseland mean opacities we may expect to get
the true result instead of just an approximate result. But in regions where
the flux is extremely small, such as in the midplane of an extremely
optically thick disc, taking the average of the opacity based on a quantity
that is nearly zero is dangerous and may lead to large errors. In spite of
these caveats the VET algorithm gives reasonable results for most problems.


\section{Summary}
We have presented solutions for the continuum radiative transfer in
high-opacity circumstellar discs. The problems have optical depths up
to $10^6$ and include anisotropic scattering. They represent realistic configurations for discs
around low mass stars and validate the
use of the codes to model current and future observations of discs.

We have compared the results of \NMCcodes\ independent Monte Carlo
codes for the temperature structure,
SEDs, scattered light images and/or polarisation maps, in the case of
anisotropic scattering. Overall, the
agreement between codes is very good. In the most optically thick case,
SEDs agree within 20\,\% over almost all of the wavelength range. Differences
become larger only at wavelengths shorter than 0.2\,$\mu$m for edge-on
configurations, \emph{i.e.} when the flux is extremely low and not
observed in practice. Pixel-to-pixel differences in high-resolution scattered light
images remain limited to 10\,\% and the polarisation maps do not
differ by more than 5 points of polarisation degree in regions where the
polarisation can be effectively measured by observations.
Each observation (SED, image or polarisation map)
was reproduced by at least three of the codes, providing robust
solutions to test other RT codes.

 The benchmark problems represent challenging test cases for RT codes. The
convergence of Monte Carlo methods alone become extremely slow for the
most optically thick cases and specific
numerical schemes are required to efficiently compute the temperature
structure, emerging SEDs and images, for instance
combining a Monte Carlo approach with ray-tracing and/or
diffusion approximation methods.

Comparisons between Monte Carlo codes and discrete ordinate codes, in the
cases with isotropic scattering or without scattering, show relatively
large differences, that increase with optical depth. Ray-tracing and VET
methods were successfully compared to Monte Carlo methods up to
moderate optical depths ($\tau_V$ = 100, \P04) but the convergence of
such codes
seem to become delicate in the test cases presented here. They
provide good approximate solutions but must be used with care
at high optical depths, when the goal is to perform detailed
comparisons with observations.

\begin{acknowledgements}
C.\,Pinte and F.\,M\'enard would like to thank G.\,Duch\^ene for
fruitful discussions on \mcfost. 
Some of the computations presented in this paper were performed at the Service
Commun de Calcul Intensif de l'Observatoire de Grenoble (SCCI) and on
the University of Exeter's SGI Altix ICE~8200 supercomputer.
C.\,Pinte acknowledges the funding from the European Commission's
Seventh Framework Program as a Marie Curie Intra-European Fellow
(PIEF-GA-2008-220891).
\end{acknowledgements}

\bibliographystyle{aa}
\bibliography{biblio}

\end{document}